\documentclass[a4paper]{revtex4}
\usepackage[letterpaper]{geometry}
\usepackage{float}
\usepackage{bm}
\usepackage{amssymb}
\usepackage{graphicx}
\usepackage{amsmath}
\usepackage{enumerate}
\usepackage{subfigure}

\begin{document}
{\large
\title{Mixture of two ultra cold bosonic atoms confined in a ring: stability and persistent currents}
\author{E. T. D. Matsushita and E. J. V. de Passos}
\affiliation{Instituto de F\'{i}sica, Universidade de S\~{a}o Paulo, Caixa Postal 66318,
C.E.P. 05389-970, S\~{a}o Paulo, S\~{a}o Paulo, Brazil}

\begin{abstract}
In this article we investigate the stability of quantized yrast (QY) states in a mixture  of two distinguishable equal mass bosonic atoms, $A$ and $B$, confined in a ring. We focus our investigation in the study of the energetic stability of the QY states since the Bloch analysis and the Bogoliubov theory establish that only energetically stable QY states are capable of sustain a persistent current. Based on physical considerations the stability is studied in two different two-dimensional planes. One is when we are studying the stability of a single QY state. In this case the study is realized in the $U_{AB}\times U$ plane spanned by the inter and intraspecies interaction strengths, $U_{AB}$ and $U$, with fixed values of angular momentum per particle $l$ and population imbalance $f$ equal to the labels of the QY state. We found that the energetic phase boundary is the positive branch of a hyperbola and the energetically stable domain the internal region of this positive branch. The other is when we are studying the stability at a fixed dynamics. In this case the stability is studied in the $l\times f$ plane spanned by the system parameters $l$ and $f$ with fixed values of the interaction strengths. In our procedure we begin considering $l$ and $f$ as continuous and unbounded quantities defined in all $l\times f$ plane and we present the stability diagram in this plane. The QY states are introduced when we postulate a correspondence between points in the $l\times f$ plane of coordinates $(l,f)$ and QY states whose labels are the coordinates of these points. As in the viewpoint of physics $f$ is a bounded quantity, this correspondence is restricted to the sector of the $l\times f$ plane of physical significance (SPS) defined by inequalities $-\infty< l<\infty$ and $-1\leq f\leq 1$. Thus what is physically significant is the stability diagram in the SPS which is determined by the overlap of the stability diagram in all $l\times f$ plane and the SPS. Loosely speaking the SPS stability diagram is distilled from the stability diagram in $l\times f$ plane. We found that there are critical values of $f$ and $l$. $f_\mathrm{crit}(l)$ gives the size of the window of energetic stability in the sense that for a given $l$ only QY states whose $f$ is in the interval $0\leq f<f_\mathrm{crit}(l)$ are energetically stable. On the other hand, $l_\mathrm{crit}$ establishes an upper bound in the angular momentum per particle of energetically stable QY state in the sense that there is none energetically stable QY state with $l>l_\mathrm{crit}$.
\end{abstract}
\maketitle
\section{ Introduction }
The properties of superfluidity of a system of ultra cold atoms confined in a ring
has been extensively studied  in recent years, both experimentally \cite{Rama,Moulder,Wright,Murray,Beattie} and 
theoretically \cite{Kana,Kavou3,Kavou4}. The experiments of references \cite{Rama,Moulder,Wright,Murray,Beattie} have managed to create in 
the laboratory persistent currents in this system.

At the theoretical side an analysis by Bloch \cite{Bloch} concluded that the occurrence of persistent currents  is related to the stability of the 
yrast states. As the angular momentum $L_z$ commutes with the Ring Hamiltonian $H$ the 
stationary states can be chosen to be  simultaneous eigenstates of $L_z$ and $H$.
The state with the lowest energy for a given $L$, where $L$ is an eigenvalue of $L_z$, is
referred to as yrast state. According to Bloch only yrast states which 
are  local minima of the yrast  spectrum are capable of sustain 
a persistent current. Variational methods \cite{Kavou3,Kavou4}, the Bogoliubov theory (when applicable) \cite{Kana} and a truncated diagonalization of the Ring Hamiltonian \cite{Kana}
have been employed to calculate the yrast spectrum.
The main conclusion from these calculations was that only yrast states
with an integer angular momentum per particle $ \frac{L}{N} =l $, with 
$l=0,1,2,\ldots$, is capable of sustain a persistent current.

A non trivial extension of the previous case is to investigate persistent currents in systems which are a
mixture of two distinguishable atoms confined in the ring. If $A$ and $B$ are 
the labels of the two species, we have a mixture of $ N_{A} $ atoms of specie $A$ 
 and $N_{B} $ atoms of specie $B$, the total number of atoms being equal to
$ N= N_{A}+ N_{B} $ and the total angular momentum equal to $ L= L_{A}+ L_{B} $. At fixed value of $N$ the number of atoms of each species can be parameterized as $N_{A}=\frac{N}{2}(1-f)$ and $N_{B}=\frac{N}{2}(1+f)$ where $f$ is the population imbalance, $-1\leq f\leq1$.
The atoms interact via a contact interaction with strengths 
$ U_{AA} , U_{BB} $ and $ U_{AB} $. Quantitative studies of this system have been done only by variational
methods in the limit of equal \cite{Smyr9,Zaremba13,Anosh,Smyr14,Abad} and unequal \cite{Zaremba15} interaction strengths.

In this article we investigate, in the framework of the Bogoliubov theory,
the stability properties of quantized yrast (QY) states in a mixture of two equal mass
distinguishable atoms. In our analysis we consider equal intraspecies interaction strengths, $U_{AA}= U_{BB}\equiv U$ different from interspecies interaction strength $U_{AB}$.  In the Bogoliubov theory the determination of the stability of an equilibrium 
state is based on the following criterion:  (i) if the energies of the 
elementary excitations are all real the equilibrium state is dynamically stable \cite{Wu,Wu_Niu,Paraoanu}; (ii) if the  energies of the elementary excitations are all real and positive the equilibrium state is energetically stable \cite{Zaremba15,Paraoanu,Baym}. This criterion is based on the following property. In the Bogoliubov theory there is a transformation that diagonalizes any quadratic form, in particular, the Effective Hamiltonian \cite{Blaizot}. The eigenvalues can be complex or real. If at least one eigenvalue is complex the equilibrium state is dynamically unstable. When the eigenvalues are real they come in pairs of eigenvectors with opposite eigenvalues and opposite norm. The elementary excitation is assigned to the eigenvector of  positive norm whose eigenvalue is the energy of the elementary excitation. If there is at least one eigenvector with positive norm and negative eigenvalue the equilibrium state is energetically unstable. If all the eigenvectors of positive norm have positive eigenvalues the equilibrium state is energetically stable \cite{Zaremba15,Paraoanu,Baym}.

Since an energetically stable QY state is a local minimum and taking into consideration the Bloch analysis \cite{Bloch} we conclude that only the energetically stable QY states are capable of sustain a persistent current. We focus our investigation in the study of the energetic stability of the QY states. We solved the Bogoliubov-de Gennes equations of the model to find analytic expressions for the energies of the elementary excitations written in terms of the system parameters $(U_{AB},U,l,f)$. These system parameters span a four-dimensional system parameters space. The QY state is specified by two labels, the angular momentum per particle $l$ and the population imbalance $f$. The dynamics is specified by $U$ and $U_{AB}$. Based on Bogoliubov criterion we determine the inequalities that when satisfied define the region of the system parameters space where the QY states are energetically stable. The inequalities when expressed in terms of the system parameters are of the form $\mathcal{F}_1(U_{AB},U,l,f)>0$ and $\mathcal{F}_2(U_{AB},U,l,f)>0$, therefore the energetically stable domain is given by the intersection of the domains of each inequality $\mathcal{F}_1\cap\mathcal{F}_2 $. The boundary of the $\mathcal{F}_1\cap\mathcal{F}_2 $ domain is an hypersurface in the system parameters space  which will be identified as the phase boundary hypersurface. It is very difficult to handle these inequalities in the four-dimensional space. Nevertheless we found two cases where physical considerations, taking into account that we are studying the stability of QY states, reduce the study in the four-dimensional space to a study in a two-dimensional plane. In the first case we are studying the stability of a single QY state. In this case the values of the system parameters $l$ and $f$ should be fixed equal to the labels of the QY state, $l=l_\mathrm{p}$ and $f=f_\mathrm{p}$. If we fix the values of $l$ and $f$ in the phase boundary hypersurface it reduces to a curve in a two-dimensional plane spanned by the system parameters $U_{AB}$ and $U$. This curve is the parameter dependent phase boundary in the $U_{AB}\times U$ plane and is the intersection of the phase boundary hypersurface with the hyperplanes  $l=l_\mathrm{p}$ and $f=f_\mathrm{p}$. The stability diagram in the $U_{AB}\times U$ plane determine the region of energetic stability in this plane. In the second case we are studying the stability at a fixed dynamics. If we fix the values of $U_{AB}$ and $U$ in the phase boundary hypersurface it reduces to a curve in a two-dimensional plane spanned by the system parameters $l$ and $f$. This curve is the parameter dependent phase boundary in the $l\times f$ plane and is equal to the intersection of the phase boundary hypersurface  with the hyperplanes $U_{AB}=U_{AB\mathrm{p}}$ and $U=U_\mathrm{p}$. The stability diagram in the $l\times f$ plane determine the region of energetic stability in this plane. The QY states appear when we postulate a correspondence between points in the $l\times f$ plane of coordinates $(l,f)$ and the QY states whose labels are the coordinates of these points. Until now we consider $l$ and $f$ as continuous and unbounded quantities. However, in the viewpoint of physics, $f$ is a bounded quantity and the correspondence is restricted to the sector of the $l\times f$ plane of physical significance (SPS) defined by inequalities $-\infty< l<\infty$ and $-1\leq f\leq 1$. This correspondence extends to the stability properties. The regime of stability of a QY state whose labels are $l$ and $f$ is equal to the regime of stability of the corresponding point in the SPS. Therefore we can identify the energetically stable QY states as the ones whose corresponding points in the SPS are localized in the energetically stable domain in the $l\times f$ plane.

Recently it appeared in the literature a theoretical work \cite{Zaremba15} which, besides other applications, investigate the stability of QY states. Both of us found identical inequalities that define the region of energetic stability of the mixture and studied the stability in planes spanned by pairs of system parameters. We differ by the choice of the planes. Our choice of the $U_{AB}\times U$ and $l\times f$ planes is based on its physical significance and straightforward physical interpretation. On the other hand, the reference \cite{Zaremba15} overlooked the importance of these planes choosing instead to study the stability, broadly speaking, in the $U\times f$ plane.
                 
This article is organized as follows: In Section II we describe the system
under consideration and we define the states that will be identified with the QY states compatible with the Bogoliubov theory. We discussed briefly our method to solve the Bogoliubov-de Gennes equations of the model and we found analytic expressions for the energy of the elementary excitations written in terms of the system parameters. Based on the Bogoliubov stability criterion, we determine the inequalities that when satisfied are the necessary and sufficient conditions for QY states to be dynamically and energetically stable. In Section III we present the stability diagram in the  $U_{AB}\times U$ plane to show that the energetic phase boundary is the positive branch of a hyperbola and the energetically stable domain is the internal region of this positive branch. The stability in the rarefied limit was also examined. In Section IV we present the stability diagram in the SPS and we investigate the consequences of the occurrence of two critical quantities: $f_\mathrm{crit}(l)$ and $l_\mathrm{crit}$.  In Section V we present a summary and our conclusions.

\section{Bogoliubov stability criterion}

\subsection{The System}
In our system, a two-component gas is confined in a tight toroidal trap of radius $R$ and cross-section $S$, the ring. The two-component gas is a mixture of $N_A$ atoms of specie $A$ and $N_B$ atoms of specie $B$ with $N=N_A+N_B$ fixed. The Ring Hamiltonian, in units of $\frac{\hbar^2}{2MR^2}$, in second quantization reads
\begin{equation}
H=\sum_{s}\sum_{m}m^2 a_{s,m}^{\dag}a_{s,m}+\frac{1}{2}\sum_{s,s^\prime}\sum_{m_i}U_{ss^\prime}a_{s,m_1}^{\dag}a_{s^\prime,m_2}^{\dag}a_{s,m_3}a_{s^\prime,m_4} \delta_{m_1+m_2,m_3+m_4}\label{ghamiltonian}
\end{equation}
where $a_{s,m}$ ($a_{s,m}^\dag$) is the bosonic annihilation (creation) operator of an atom of specie $s=A,B$ in an eigenstate of $l_{z}$ with eigenvalue $m$ and $U_{ss^\prime}=\frac{4Ra_{ss^\prime}}{S}$ is the interaction strength between atoms of species $s$ and $s^\prime$, in units of $\frac{\hbar^2}{2MR^2}$, with $a_{ss^\prime}$ being the respective $s-$wave scattering length. 

In this paper we investigate the stability properties of the QY states in a mixture of population imbalance $f=\frac{N_B-N_A}{N}$. In a mean field theory, a QY state of angular momentum 
equal  to $L=Nl$ is a Bose condensed state where the $N_{A}$ atoms of 
specie $A$ and  the $N_{B}$ atoms of specie $B$ occupy the same eigenstate of $l_{z}$
of eigenvalue $l$ and wave function $\phi_{l}(\theta) =\frac{1}{\sqrt{2\pi}}e^{il\theta}$.
This Bose condensed state has angular moment equal to $ L=N_{A} l+N_{B}l=Nl$
and is a uniform density solution  of the GP equations, with chemical
 potentials equal to
 \begin{equation}
\mu_{A}= l^{2}+(N_A-1) U_{AA}+N_B U_{AB} \;\;\;\;\;\textrm{and}\;\;\;\;\;\mu_{B}= l^{2}+(N_B-1) U_{BB}+N_A U_{AB}. 
\end{equation}

\subsection{Elementary excitations}
In the Bogoliubov theory the QY state is identified with the vacuum of the shifted operators $c_{s,m}$ defined by
\begin{equation}a_{s,m}=c_{s,m}+z_{s,l}\delta_{m,l}\label{shifted}\end{equation}
where $z_{s,l}$ are $c-$numbers which appear as a shift in the Bose condensed state $a_{s,l}$ and will be determined later on. Next we write the Grand-Hamiltonian $\mathcal{H}\equiv H-\sum_{s}\mu_sN_s$ as a normal order expansion with respect to the shifted operators, $\mathcal{H}=\sum_{i=0}^{4}\mathcal{H}_i$, where the term $\mathcal{H}_i$ involves $i$ shifted operators. 

The expectation value of the Grand-Hamiltonian in the vacuum state is equal to,
\begin{equation}
\langle \mathcal{H}\rangle=\sum_{s}(l^2-\mu_s)|z_{s,l}|^2+\frac{1}{2}\sum_{s,s^\prime}U_{ss^\prime}|z_{s,l}|^2|z_{s^\prime l}|^2.
\end{equation}
The vacuum is identified with the stationary state of the Grand-Hamiltonian subject to the number-conserving constraints $N_s=|z_{s,l}|^2$ for $s=A,B$ leading to the equations
\begin{subequations}
\begin{eqnarray}
\left(l^2-\mu_A+U_{AA}|z_{A,l}|^2+U_{AB}|z_{B,l}|^2\right)z_{A,l}&=&0\label{GP1}\\
\left(l^2-\mu_B+U_{BB}|z_{B,l}|^2+U_{AB}|z_{A,l}|^2\right)z_{B,l}&=&0\label{GP2}
\end{eqnarray}
\end{subequations}
The Eqs. (\ref{GP1}) and (\ref{GP2}) are invariant by a phase change of the $z_{s,l}$ therefore we can take these $c-$numbers as real and equal to $z_{A,l}=\sqrt{N_A}$ and $z_{B,l}=\sqrt{N_B}$ which reduces the Eqs. (\ref{GP1}) and (\ref{GP2}) to
\begin{subequations}
\begin{equation}
\mu_A=l^2+N_{A}U_{AA}+N_{B}U_{AB}
\end{equation}
\begin{equation}
\mu_B=l^2+N_{B}U_{BB}+N_{A}U_{AB}.
\end{equation}
\end{subequations}
This vacuum has the property of making the linear term $\mathcal{H}_1$ identically zero.

The dynamics in the neighborhood of an equilibrium state is described by an Effective Hamiltonian which is the normal ordered expansion of the Grand-Hamiltonian up to the second order in the shifted operators. Since $\mathcal{H}_0$ is a constant and $\mathcal{H}_1$ identically zero, the Effective Hamiltonian reduces to the quadratic term $\mathcal{H}_2$ given by

\begin{eqnarray}
\mathcal{H}_{2}&=&\sum_{s,s^\prime}\sum_m \left[(m^2-l^2)\delta_{s,s^\prime}+U_{ss^\prime}\sqrt{N_sN_{s^\prime}}\right]c_{s,m}^{\dag}c_{s^\prime,m}\nonumber\\&&+\frac{1}{2}\sum_{s,s^\prime}\sum_q U_{ss^\prime}\sqrt{N_sN_{s^\prime}}(c_{s,l+q}c_{s^\prime,l-q}+c_{s^\prime,l-q}^{\dag}c_{s,l+q}^{\dag}) \label{eq:effective}
\end{eqnarray}
with $q$ being the transferred angular momentum. The diagonalization of a quadratic form of bosonic operators is a standard problem fully explained in reference [17]. The energies and the composition of the elementary excitations are found solving the Bogoliubov-de Gennes (BdG) equations. To deduce the BdG equations, notice that an inspection of (\ref{eq:effective}) reveals that the coupling structure is such that only pairs $(l\pm q)_A$ and $(l\pm q)_B$ with $q\geq 0$ are coupled which allow us to express $\mathcal{H}_2$ as
\begin{subequations}
\begin{eqnarray}
\mathcal{H}_{2}&=&\sum_{s,s^\prime} h_{s,l;s^\prime,l}c_{s,l}^{\dag}c_{s^\prime,l}+\frac{1}{2}\Delta_{s,l;s^\prime,l}\left(c_{s,l}c_{s^\prime,l}+c_{s^\prime,l}^{\dag}c_{s,l}^\dag\right)\nonumber\\&+&\sum_{q>0}\sum_{\substack{s,s^\prime\\\lambda,\lambda^\prime}}\left[h_{s,l+\lambda q;s^\prime,l+\lambda^\prime q}c_{s,l+\lambda q}^{\dag}c_{s^\prime,l+\lambda^\prime q}+\frac{1}{2}\Delta_{s,l+\lambda q;s^\prime,l+\lambda^\prime q}\left(c_{s,l+\lambda q}c_{s^\prime,l+\lambda^\prime q}+c_{s^\prime,l+\lambda^\prime q}^{\dag}c_{s,l+\lambda q}^\dag\right)\right]\nonumber\\\label{eq:hdelta}
\end{eqnarray}
where $\lambda,\lambda^\prime=\pm 1$ give the sign of the transferred angular momentum $|q|$ and
\begin{eqnarray}
h_{s,l+\lambda q;s^\prime,l+\lambda^\prime q}&=&\left[(q^2+2\lambda ql)\delta_{s,s^\prime}+U_{ss^\prime}\sqrt{N_sN_{s^\prime}}\right ]\delta_{\lambda,\lambda^\prime}\label{eq:ele1}\\
\Delta_{s,l+\lambda q;s^\prime,l+\lambda^\prime q}&=&U_{ss^\prime}\sqrt{N_sN_{s^\prime}}\delta_{\lambda,-\lambda^\prime}.\label{eq:ele2}
\end{eqnarray}
The BdG equations can be seen as a non-hermitian eigenvalue problem ${\cal M} V=E\eta V $ with ${\cal M}=\left[\begin{array}{cc}h&\Delta\\\Delta^{\star}&h^{\star}\end{array}\right]$ being the BdG hermitian matrix whose elements are given by Eqs. (\ref{eq:ele1}) and (\ref{eq:ele2}), $V=\left[\begin{array}{cc} u&v\end{array}\right]^\textrm{T}$ is the eigenvector corresponding to the eigenvalue (excitation energy) $E$ and $\eta=\mathrm{diag}(1,-1)$ is the bosonic metric. 
\end{subequations}

It is known that the excitation energies for $l\neq 0$ case differ from excitation energies for $l=0$ case from a $l-$dependent shift $\pm 2ql$, \cite{Zaremba15}. In turn, in the $l=0$ case the BdG equations can be solved leading to a double degenerate spin and density modes [14] whose excitation energies are
\begin{subequations}
\begin{equation}
E_d=\sqrt{\frac{1}{2}\left[c_{AA}+c_{BB}+\sqrt{(c_{AA}+c_{BB})^2-4(c_{AA}c_{BB}-c_{AB}^2)}\right]}\label{eq:density_mode}
\end{equation}
\begin{equation}
E_s=\sqrt{\frac{1}{2}\left[c_{AA}+c_{BB}-\sqrt{(c_{AA}+c_{BB})^2-4(c_{AA}c_{BB}-c_{AB}^2)}\right]}\label{eq:spin_mode}
\end{equation}
where
\begin{equation}
c_{AA}\equiv q^2(q^2+2U_{AA}N_A),\;\;\;\;\;c_{BB}\equiv q^2(q^2+2U_{BB}N_B)\;\;\;\;\;\textrm{and}\;\;\;\;\;c_{AB}\equiv 2q^2U_{AB}\sqrt{N_AN_B}.
\end{equation}
\end{subequations}
Thus the excitation energies for $l\neq 0$ case are non-degenerate and given by $E_d\pm 2ql$ and $E_s\pm 2ql$. What is lacking is a proof of the connection between the excitation energies of $l=0$ and $l\neq 0$. An outline of a proof goes as follow. As a consequence of the coupling structure of the Effective Hamiltonian, the BdG hermitian matrix ${\cal M}$ is diagonal in blocks of dimension $4\times 4$ when $q=0$ and $8\times 8$ when $q>0$ revealing that the excitation spectrum is composed of one doublet and quadruplets specified by the value of the magnitude of the transferred angular momentum $q>0$. The doublet diagonalization gives two zero energy modes which leads to an indifferent equilibrium which does not affect the stability of the QY states and will be ignored from now on. The $8\times 8$ eigenvalue problem of BdG decomposes into two $4\times 4$ ones which when written as an eigenvalue problem for shifted eigenvalues $+2ql$ and $-2ql$ become independent of $l$ and equal to the $l=0$ case.
\subsection{Dynamical stability criterion}
According to the Bogoliubov theory an equilibrium state is dynamically stable if all the excitation energies are real. The existence of at least one complex energy is sufficient to guarantee the dynamical instability of the corresponding equilibrium state. As $E_s<E_d$ we see that the energies of the quadruplet  with transferred angular momentum $q$ are real if $E_s^2$ is real and positive. From (\ref{eq:density_mode}) and (\ref{eq:spin_mode}), this condition is satisfied if
\begin{equation}c_{AA}+c_{BB}>0,\;\;\;\;\;c_{AA}c_{BB}-c_{AB}^2>0\;\;\;\;\;\textrm{and}\;\;\;\;\;(c_{AA}+c_{BB})^2-4(c_{AA}c_{BB}-c_{AB}^2)>0.\label{eq:dsta_prim}
\end{equation}
The last inequality is always satisfied since it is equal to $(c_{AA}-c_{BB})^2+4c_{AB}^2>0$. The other two inequalities of (\ref{eq:dsta_prim}) can be cast into the form
\begin{equation}
\begin{array}{cc}(D1)&\left(U_{AA}N_{A}+\dfrac{q^2}{2}\right)+\left(U_{BB}N_{B}+\dfrac{q^2}{2}\right)>0\\ &\\(D2)&\left(U_{AA}N_{A}+\dfrac{q^2}{2}\right)\left(U_{BB}N_{B}+\dfrac{q^2}{2}\right)-U_{AB}^2N_{A}N_{B}>0\end{array}
\end{equation}
In the $U_{AA}N_{A}\times U_{BB}N_{B}$ plane these inequalities define a domain where the boundary is the positive branch of the hyperbola $\left(U_{AA}N_{A}+\frac{q^2}{2}\right)\left(U_{BB}N_{B}+\frac{q^2}{2}\right)-U_{AB}^2N_{A}N_{B}=0$ of center at $\left(-\frac{q^2}{2},-\frac{q^2}{2}\right)$ and semi-major axis $|U_{AB}|\sqrt{2N_AN_B}$. In this domain the energies of quadruplet $q$ are all real. However we need to determine the domain where the energies of all quadruplets are real. A consequence of the properties of the hyperbola is that the variation of $q$ leads to a translation of this curve along its axis $U_{AA}N_{A}-U_{BB}N_{B}=0$. As the center is a decreased function of $q$ we see that the domain of the quadruplet $q_\mathrm{min}=1$ is contained in the domain of all the other quadruplets.  Therefore, we conclude that the energies of all quadruplets are real if the energies of the quadruplet $q_\mathrm{min}=1$ are real. Thus the equilibrium state is dynamically stable if
\begin{equation}
\begin{array}{cc}(D1)&\left(U_{AA}N_{A}+\dfrac{1}{2}\right)+\left(U_{BB}N_{B}+\dfrac{1}{2}\right)>0\\ &\\(D2)&\left(U_{AA}N_{A}+\dfrac{1}{2}\right)\left(U_{BB}N_{B}+\dfrac{1}{2}\right)-U_{AB}^2N_{A}N_{B}>0\end{array}\label{eq:dyn_dyn}
\end{equation}
These conditions do not depend on the angular momentum per particle $l$ of the QY state.
\subsection{Energetic stability criterion}
According to the Bogoliubov theory an equilibrium state is energetically stable if all the excitation energies are real and positive. As $E_d>E_s$, the energies of a quadruplet $q$ are positive if $E_s>2ql$ which implies that
\begin{equation}c_{AA}+c_{BB}-8q^2l^2>0\;\;\;\;\;\textrm{and}\;\;\;\;\;(c_{AA}-4q^2l^2)(c_{BB}-4q^2l^2)-c_{AB}^2>0.\label{eq:dene_prim}
\end{equation}
The inequalities (\ref{eq:dene_prim}) can be cast into the form
\begin{equation}
\begin{array}{cc}(E1)&\left[U_{AA}N_{A}-\dfrac{4l^2-q^2}{2}\right]+\left[U_{BB}N_{B}-\dfrac{4l^2-q^2}{2}\right]>0\\ &\\(E2)&\left[U_{AA}N_{A}-\dfrac{4l^2-q^2}{2}\right]\left[U_{BB}N_{B}-\dfrac{4l^2-q^2}{2}\right]-U_{AB}^2N_{A}N_{B}>0\end{array}
\end{equation}
The analysis of these inequalities is analogous to the previous case. The only change is that the center of the hyperbola is now $\left(\frac{4l^2-q^2}{2},\frac{4l^2-q^2}{2}\right)$. It follows that if the energies of the quadruplet $q_\mathrm{min}=1$ are positive then the energies of all quadruplets are positive. Thus the equilibrium state is energetically stable if
\begin{equation}
\begin{array}{cc}(E1)&\left[U_{AA}N_{A}-\dfrac{4l^2-1}{2}\right]+\left[U_{BB}N_{B}-\dfrac{4l^2-1}{2}\right]>0\\ &\\(E2)&\left[U_{AA}N_{A}-\dfrac{4l^2-1}{2}\right]\left[U_{BB}N_{B}-\dfrac{4l^2-1}{2}\right]-U_{AB}^2N_{A}N_{B}>0\end{array}\label{eq:ene_ene}
\end{equation}
Different from the dynamical stability conditions, the energetic stability conditions depend on the angular momentum per particle $l$ of the QY state.

\section{Stability in the $U_{AB}\times U$ plane}
In this section we are interested in the study of the properties of energetic stability of a single QY state of angular momentum per particle equal to $l$ in a mixture of 
population imbalance equal to $f$, in function of the intraspecies, $u$, and interspecies, $u_{AB}$, interaction strengths where $u\equiv NU$ and $u_{AB}\equiv NU_{AB}$. In the discussion that follows we will not consider the $l=0$ state since it does not carry a current. Besides in a $l=0$ state, energetic stability is equivalent to dynamical stability.

\subsection{Energetic Stability}
 In terms of the system parameters the inequalities (\ref{eq:ene_ene}) take the form
\begin{equation}
\begin{array}{cc}(E1)& u-(4l^2-1)>0\\ &\\(E2)&(4l^2-1)^2-2 (4l^2-1)u+(1-f^{2})( u^{2} -u_{AB}^{2}) >0\end{array}\label{eq:ene_ene1}
\end{equation}
They are the inequalities that when satisfied define the region of the four-dimensional system parameters space where QY state is energetically stable. These inequalities can be expressed as
\begin{equation}
\begin{array}{cc}(E1)& u-(4l^2-1)>0\\ &\\(E2)&\left[u -\frac{4l^2-1 }{1-f^{2}}\right]^{2} - u_{AB}^{2}-\left[ \frac{ (4l^2-1) f}
{1-f^{2}}\right]^{2}>0\end{array}\label{eq:ene_ene2}
\end{equation}
Fixed the values of $l$ and $f$ equal to the labels of the QY states, $l=l_\mathrm{p}$ and $f=f_\mathrm{p}$, (\ref{eq:ene_ene2}) reduces to
\begin{equation}
\begin{array}{cc}(E1)& u-(4l_\mathrm{p}^2-1)>0\\ &\\(E2)&\left[u -\frac{4l_\mathrm{p}^2-1 }{1-f_\mathrm{p}^{2}}\right]^{2} - u_{AB}^{2}-\left[ \frac{ (4l_\mathrm{p}^2-1) f_\mathrm{p}}
{1-f_\mathrm{p}^{2}}\right]^{2}>0\end{array}\label{eq:ene_ene3}
\end{equation}
The next step is to determine the domains of these two inequalities. The inequality (E2) is a second order polynomial in $u$ of roots equal to 
\begin{equation}
u_{\pm}=\frac{4l_\mathrm{p}^2-1}{1-f_\mathrm{p}^{2}} \pm\sqrt{u_{AB}^2+a^2(l_\mathrm{p},f_\mathrm{p})}\label{eq:emerson0}
\end{equation}
where
\begin{equation} 
a(l_\mathrm{p},f_\mathrm{p})=\left|\frac{ (4l_\mathrm{p}^2-1)f_\mathrm{p}}{1-f_\mathrm{p}^{2}}\right|.\label{eq:semimajor}
\end{equation}
The curves $u=u_{\pm}$ are, respectively, the positive and negative branches of the hyperbola 
\begin{equation}
\left[u -\frac{4l_\mathrm{p}^2-1 }{1-f_\mathrm{p}^{2}}\right]^{2} - u_{AB}^{2}-\left[ \frac{ (4l_\mathrm{p}^2-1) f_\mathrm{p}}
{1-f_\mathrm{p}^{2}}\right]^{2}=0\label{eq:hyperbolaA}
\end{equation}
whose center is at $ \left(0,\frac{4l_\mathrm{p}^2-1}{1- f_\mathrm{p}^{2}}\right) $ and semi-major axis equal to $a(l_\mathrm{p},f_\mathrm{p})$. The positive branch has vertice at $ \left(0,\frac{4l_\mathrm{p}^2-1}{1- f_\mathrm{p} }\right) $ and domain $u>\frac{4l_\mathrm{p}^2-1}{1- f_\mathrm{p}}$ whereas the negative branch has vertice at $ \left(0,\frac{4l_\mathrm{p}^2-1}{1+f_\mathrm{p} }\right) $ and domain $u<\frac{4l_\mathrm{p}^2-1}{1+f_\mathrm{p} }$. The inequality (E2) can be written as $(u-u_+)(u-u_-)>0$ which is equivalent to (a) $u-u_+>0$ and $u-u_->0$ or (b) $u-u_+<0$ and $u-u_-<0$. Since $u_-<u_+$, it is easily seen that they reduce to (a) $u-u_+>0$ and (b) $u-u_-<0$. The boundaries of inequalities $u>u_+$ and $u<u_-$ are, respectively, $u=u_+$, the positive branch, and $u=u_-$, the negative branch of the hyperbola (\ref{eq:hyperbolaA}). The domain of (E2) is the internal region of the positive branch if $u>\frac{4l_\mathrm{p}^2-1}{1- f_\mathrm{p}}$ and the internal region of the negative branch if $u<\frac{4l_\mathrm{p}^2-1}{1+f_\mathrm{p} }$ with a gap in the $u$ axis in the interval $\frac{4l_\mathrm{p}^2-1}{1+f_\mathrm{p} }<u<\frac{4l_\mathrm{p}^2-1}{1-f_\mathrm{p} }$. Turning to  inequality (E1) we see that the boundary is the straight line $u=4l_\mathrm{p}^2-1$ and the domain is the semi-plane $u>4l_\mathrm{p}^2-1$. 

Now we should determine the intersection between (E1) and (E2). As $\frac{4l_\mathrm{p}^2-1}{1+f_\mathrm{p} }<4l_\mathrm{p}^2-1<\frac{4l_\mathrm{p}^2-1}{1-f_\mathrm{p} }$ we see that the boundary of (E1) crosses the $u$ axis in the gap region. As the region in the domain of (E1) is always greater than $4l_\mathrm{p}^2-1$ there is no overlap between the domains of (E1) and of the negative branch therefore the later is discarded. To conclude the energetic phase boundary is the positive branch of the hyperbola (\ref{eq:hyperbolaA}) where the energetically stable domain is the internal region of this positive branch, 
\begin{equation}
u>\frac{4l_\mathrm{p}^2-1}{1-f_\mathrm{p}^{2}} +\sqrt{u_{AB}^2+a^2(l_\mathrm{p},f_\mathrm{p})}.\label{eq:emerson}
\end{equation}

\subsection{Energetic stability and the persistent currents}
Our calculation of the stability diagrams in the $u_{AB}\times u$ plane for different values of $l_\mathrm{p}$ and $f_\mathrm{p}$ is shown in Fig.1. We select a mixture of 
equal population  $f_\mathrm{p}=0\;(0.50, 0.50 )$, one of  moderate imbalance $f_\mathrm{p}=0.50\;(0.25,0.75)$
and one of rarefied minority component  $f_\mathrm{p}=0.98\;(0.01,0.99)$ where $(n_A,n_B)$ are the fractions of atoms of each specie. For each value of $f_\mathrm{p}$
we consider $l_\mathrm{p}=1,2,3$. The light gray area is the region in the $u_{AB}\times u$ plane where the QY state is energetically stable, consequently, capable to sustain a persistent current. The orange curve is the phase boundary in the $u_{AB}\times u$ plane which is the positive branch of hyperbola $u=u_+$ with $u_+$ given in Eq. (\ref{eq:emerson0}).

\begin{figure}[h]
\centering
\includegraphics[scale=0.7]{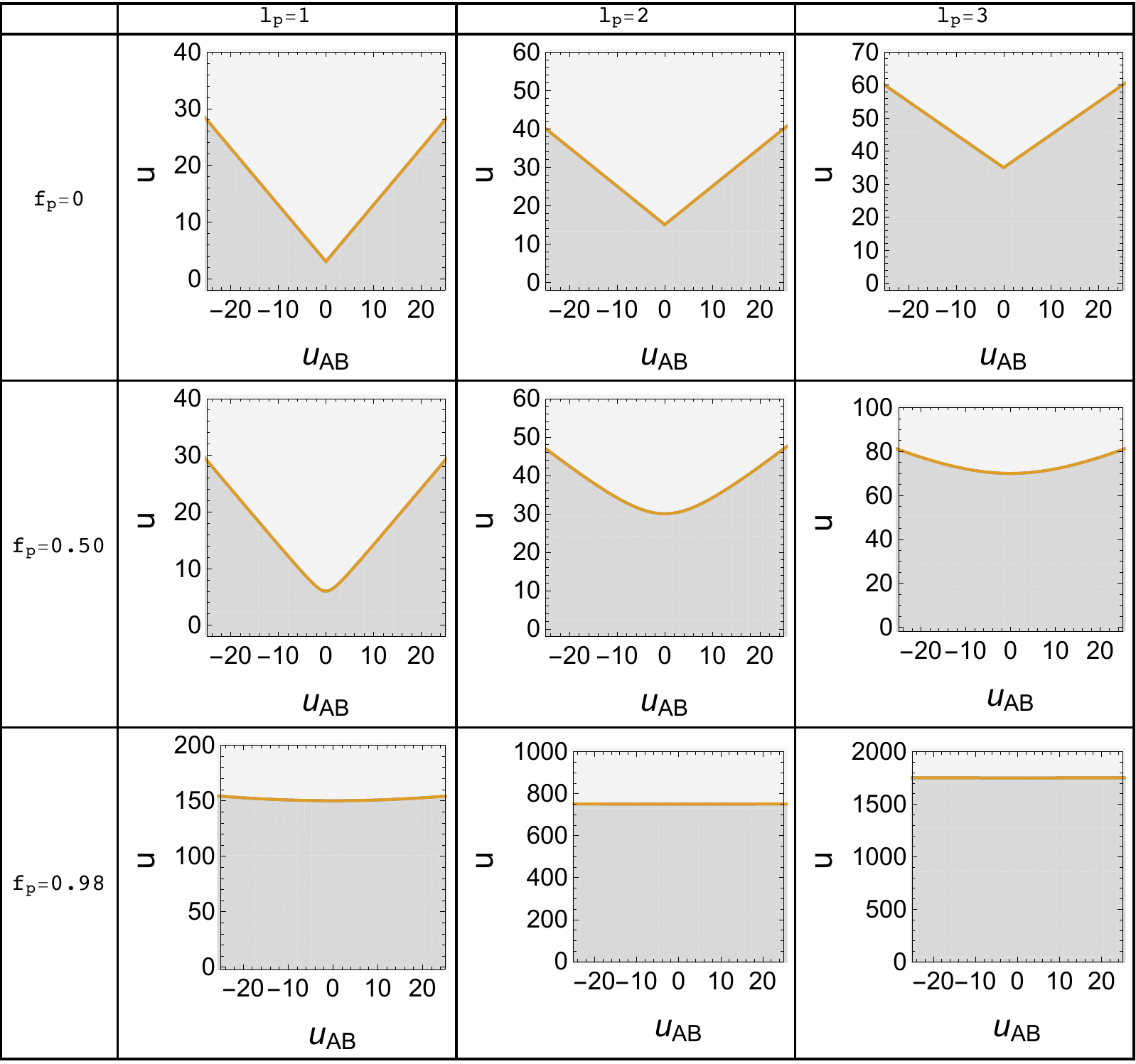}
\caption{(Color online) Stability diagrams in the $u_{AB}\times u$ plane for different values of population imbalance $f_\mathrm{p}$ (rows) and angular momentum per particle $l_\mathrm{p}$ (columns). Each pair of values $l_\mathrm{p}$ and $f_\mathrm{p}$ selects a QY state. The light gray area is the region in the $u_{AB}\times u$ plane where the QY state is energetically stable, consequently, capable to sustain persistent currents. The orange curve is the phase boundary in the $u_{AB}\times u$ plane which is the positive branch of the hyperbola (\ref{eq:hyperbolaA}). These graphs exhibit the dependence of the phase boundary in the $u_{AB}\times u$ plane on the parameters $f_\mathrm{p}$ and $l_\mathrm{p}$. Indeed when $f_\mathrm{p}$ starts to increase, for a fixed value of $l_\mathrm{p}$, the region where the phase boundary is nearly constant increases and in the limit $f_\mathrm{p}\rightarrow 1$ the phase boundary is a constant equal to $u_\mathrm{min}(l_\mathrm{p},f_\mathrm{p})=\frac{4l_\mathrm{p}^2-1}{1-f_\mathrm{p}}$. Notice that $u=u_+$ grows very fast when $f_\mathrm{p}$ approaches 1.}
\label{Rotulo}
\end{figure}

The phase boundary in the $u_{AB}\times u$ plane depends on the parameters $l_\mathrm{p}$ and $f_\mathrm{p}$ through the dependence of the hyperbola parameters on these quantities. In Fig.1 it is displayed the graphs of the phase boundaries with different values of $f_\mathrm{p}$ (rows in Fig.1) and $l_\mathrm{p}$ (columns in Fig.1) which exhibit the dependence of this curve on these quantities. Our interest is to establish how the phase boundary in the $u_{AB}\times u$ plane (orange curve in Fig.1) changes when $f_\mathrm{p}$ varies from 0 to 1. For $f_\mathrm{p}=0$ the phase boundary are the straight lines $u=4l_\mathrm{p}^2-1+|u_{AB}|$ which coincides with the asymptotes of the hyperbola. For $f_\mathrm{p}\neq 0$, $u$ is an increased function of $|u_{AB}|$ with minimum at $u_{AB}=0$, the value of $u$ at this minimum equal to $u_\mathrm{min}(l_\mathrm{p},f_\mathrm{p})=\frac{4l_\mathrm{p}^2-1}{1-f_\mathrm{p}}$. When $f_\mathrm{p}$ starts to increase it appears a range of values of $u_{AB}$ for which $u$ is nearly a constant equal to $u_\mathrm{min}(l_\mathrm{p},f_\mathrm{p})$. This range of values of $u_{AB}$ defines a domain in the $u_{AB}$ axis that we called the lower region. The lower region is defined by all values of $u_{AB}$ satisfying the inequality $\frac{|u_{AB}|}{a(l_\mathrm{p},f_\mathrm{p})}\ll 1$. To leading order, the phase boundary $u=u_+$ has identical values equal to  $u_\mathrm{min}(l_\mathrm{p},f_\mathrm{p})$ at any $u_{AB}$ in the lower region. If $f_\mathrm{p}$ continuous to increase, $a(l_\mathrm{p},f_\mathrm{p})$ increases and the lower region grows and in the limit $f_\mathrm{p}\rightarrow 1$ ($a\rightarrow \infty$) the domain of the lower region is all the $u_{AB}$ axis, $u$ is independent of $u_{AB}$ and equal to $u_\mathrm{min}(l_\mathrm{p},f_\mathrm{p})$. The exact calculations displayed in Fig. 1 confirm the above behavior.

To continue our discussion we will investigate the $f_\mathrm{p}\rightarrow 1$ limit of the mixture in a different context. In the previous discussion, it was shown that the $f_\mathrm{p}\rightarrow 1$ limit of the phase boundary in the $u_{AB}\times u$ plane is independent of $u_{AB}$. This suggests the interpretation that, in this limit, we have a mixture of two non-mutually interacting gases: a majority component and a rarefied minority component. Of course the condition of energetic stability of the mixture reduces to conditions of energetic stability for each component (\ref{eq:ene_ene}) (a) $u>\frac{4l_\mathrm{p}^2-1}{2n_A}$ and (b) $u>\frac{4l_\mathrm{p}^2-1}{2n_B}$. As $n_A\ll n_B$ we see that (b) is automatically satisfied if (a) is satisfied. Therefore the condition of energetic stability is given by $u>\frac{4l_\mathrm{p}^2-1}{1-f_\mathrm{p}}$ since $n_A=\frac{1-f_\mathrm{p}}{2}$. Note that the boundary $u=\frac{4l_\mathrm{p}^2-1}{1-f_\mathrm{p}}$ coincides with the $f_\mathrm{p}\rightarrow 1$ limit of the phase boundary in the $u_{AB}\times u$ plane. Therefore we conclude that the stability of the mixture when $f_\mathrm{p}\rightarrow 1$ is dominated by the stability of the minority component.

\section{Stability in the $l\times f$ plane}
Here we adopt the standard procedure of consider $l$ and $f$ as continuous and unbounded quantities. However the physical values of $l$ and $f$ are restricted to the sector of $l\times f$ plane of physical significance (SPS) defined by the inequalities $-\infty<l<\infty$ and $-1\leq f\leq 1$. In our study of energetic stability in the $l\times f$ plane we found useful to extend the discussion to include dynamical stability. In this section, our task in this section is mainly to determine for fixed values of $u$ and $u_{AB}$, $u=u_\mathrm{p}$ and $u_{AB}=u_{AB\mathrm{p}}$,  which are the energetically stable QY states.

\subsection{Dynamical Stability}
The inequalities (\ref{eq:dyn_dyn}) can be written as
\begin{equation}
\begin{array}{cc}(D1)& u_\mathrm{p}>-1\\ &\\(D2)&f^2<p(0)\end{array}\label{eq:dyn_ene1}
\end{equation}
where $p(0)$ is the polynomial $p(l)$,
\begin{equation}
p(l)=\dfrac{(4l^2-1-u_\mathrm{p})^2-u_{AB\mathrm{p}}^2}{u_\mathrm{p}^2-u_{AB\mathrm{p}}^2}, \label{eq:pl}
\end{equation}
evaluated at $l=0$, where we already fix the values of the interaction strengths. The inequalities (D1) and (D2) are equivalent to
\begin{equation}
\begin{array}{cc}(D1)& u_\mathrm{p}>-1\\ &\\(D2)&-\sqrt{p(0)}<f<\sqrt{p(0)}\end{array}\label{eq:dyn_ene2}
\end{equation}
Thus we can assert that the dynamically stable domain in the $l\times f$ plane is the stripe $-\sqrt{p(0)}<f<\sqrt{p(0)}$. As $p(0)$ is always greater than 1 the SPS is immersed in the dynamically stable domain. This lead us to conclude that all the QY states are dynamically stable, see Fig. (2a). 

\subsection{Energetic stability}
The dynamically stable domain can be split as the union of two disjoint domains: the energetically stable and unstable domains. Besides we can define two limits: one when dynamical stability is equivalent to energetic stability and the other when dynamical stability is equivalent to energetic instability. Having said this, to start  our discussion we write the inequalities (E1) and (E2) in the form
\begin{equation}
\begin{array}{cc}(E1)&l^2<\dfrac{u_\mathrm{p}+1}{4}\\ &\\(E2)& f^2<p(l)\end{array}
\end{equation}
with $p(l)$ given in Eq. (\ref{eq:pl}). We begin the analysis by the inequality (E2). As $f^2$ is a positive quantity, $p(l)$ cannot be negative which requires that we determine its sign. $p(l)$ is a quadratic polynomial in $l^2$ whose positive roots are
\begin{equation}
l_{+}^2=\frac{1}{4}(u_\mathrm{p}+1+|u_{AB\mathrm{p}}|)\;\;\;\;\mathrm{and}\;\;\;\; l_{-}^2=\frac{1}{4}(u_\mathrm{p}+1-|u_{AB\mathrm{p}}|).
\end{equation}
It is well known that once we know the roots the sign of the polynomial are easily determined: (a) $p(l)>0$ if (i) $l^2>l_{+}^2$ or (ii) $l^2<l_{-}^2$; (b) $p(l)<0$ if $l_{-}^2<l^2<l_{+}^2$. The inequalities (a) are equivalent to (i) $l>l_{+}$ or $l<-l_{+}$ and 
(ii) $-l_{-}<l<l_{-}$. On the other hand the case (b) has no physical significance and it will be discard. This implies that the positive values of $p(l)$ have gaps in the intervals $-l_{+}<l<-l_{-}$ and $l_{-}<l<l_{+}$. 

In the region where $p(l)>0$ we can write the inequality (E2) as $(f+\sqrt{p(l)})(f-\sqrt{p(l)})<0$ which is equivalent to (a) $f<\sqrt{p(l)}$ and $f>-\sqrt{p(l)}$ or (b) $f>\sqrt{p(l)}$ and  $f<-\sqrt{p(l)}$. The two inequalities in (b) are incompatible leaving only the inequalities in (a). At this stage we point out that inequalities (E1) and (E2) are equivalent to
\begin{equation}
\begin{array}{cc}(E1)&l^2<\dfrac{u_\mathrm{p}+1}{4}\\ &\\(E2)& -\sqrt{p(l)}<f<\sqrt{p(l)}\end{array}
\end{equation}
where we see that (E2) is the intersection of the inequalities $f<\sqrt{p(l)}$ and $f>-\sqrt{p(l)}$. The boundary of inequality $f<\sqrt{p(l)}$ is the curve $f=\sqrt{p(l)}$. In the interval $-l_{-}<l<l_-$, $f$ vanishes at the extrema, $f(\pm l_{-})=0$, and for $l$ positive (negative) it is a decreased (increased) function of $l$ with a maximum at $l=0$ and the value of $f$ at the maximum always greater than 1 and equal to $p(0)$. In the intervals $-l_{+}<l<-l_{-}$ and $l_-<l<l_+$ there is a gap and, at last, in the intervals $l>l_+$ and $l<-l_+$, $f$ vanishes at the extrema, $f(\pm l_{+})=0$, and it is a (increased) decreased function of $l$. However they cannot increase (decrease) indefinitely since they hit the boundary of dynamical stability at the points $l=\pm \sqrt{\frac{u_\mathrm{p}^2+1}{2}}$. The boundary of inequality $f>-\sqrt{p(l)}$ is the curve $f=-\sqrt{p(l)}$ which is the reflection with respect to the $l$ axis of the boundary $f=\sqrt{p(l)}$. The domain of (E2), $f^2-p(l)<0$, is the intersection of these two domains. Turning to (E1) we can show that its domain is the stripe $-\sqrt{\frac{u_\mathrm{p}+1}{4}}<l<\sqrt{\frac{u_\mathrm{p}+1}{4}}$ where the boundaries are the straight lines $l=\pm \sqrt{\frac{u_\mathrm{p}+1}{4}}$.

To determine the intersection of (E1) and (E2) notice that as $l_-<\sqrt{\frac{u_\mathrm{p}+1}{4}}<l_+$, the boundaries of (E1) cross the $l$ axis in the gap regions. As the domain of (E1) is the internal region of the stripe we see that (E1) discards the domains in the $l<-l_{+}$ and $l>l_{+}$ regions leaving only the domain in the $-l_{-}<l<l_{-}$ region. Thus the domain of energetic stability given by the intersection of (E1) and (E2) is specified by the inequalities $-l_\mathrm{crit}<l<l_\mathrm{crit}$ and $-f_\mathrm{crit}(l)<f<f_\mathrm{crit}(l)$ where $l_\mathrm{crit}= l_- $ and $f_\mathrm{crit}(l)= \sqrt{p(l)}$ are equal to
\begin{equation}
l_\mathrm{crit}=\frac{1}{2}\sqrt{u_\mathrm{p}+1-|u_{AB\mathrm{p}}|} \;\;\;\;\;\mathrm{and}\;\;\;\;\; f_\mathrm{crit}(l)= \sqrt{\frac{(4l^2-1-u_\mathrm{p})^2-u_{AB\mathrm{p}}^2}{u_\mathrm{p}^2-u_{AB\mathrm{p}}^2}}
\end{equation}
The phase boundary is the closed curve $f=f_\mathrm{crit}(l)$ if $-l_-<l<l_-$ and $0<f<\sqrt{p(0)}$ and $f=-f_\mathrm{crit}(l)$
if $-l_-<l<l_-$ and $-\sqrt{p(0)}<f<0$ where the energetic stability domain is the internal region of this closed curve. In Fig. (2a) we present the stability diagram in the $l\times f$ plane.

Until now we consider $l$ and $f$ as continuous and unbounded quantities defined in all $l\times f$ plane. However from the viewpoint of the physics, the correspondence between points of coordinates $(l,f)$ and QY states is restricted to the SPS. Thus, what is physically significant is to present the stability diagram in the SPS which is determined by the overlap of the SPS and the stability diagram in all $l\times f$ plane. As a consequence of this overlap the energetic phase boundary is now the curve
\begin{equation}
f=\begin{cases} \pm f_\mathrm{crit}(l)&\textrm{if }-l_-<l<-\frac{1}{2}\\\pm1&\textrm{if }-\frac{1}{2}<l<\frac{1}{2}\\\pm f_\mathrm{crit}(l)&\textrm{if }\frac{1}{2}<l<l_- \end{cases}
\end{equation}
and the energetically stable domain is the internal region of this curve. Notice that the curve for the negative values of $f$ are the reflection with respect to the $l$ axis of the curve for the positive values, see Fig. (2b) and (2c). The energetically stable QY states are those whose corresponding points in the SPS are in the energetically stable domain in the $l\times f$ plane.

\begin{figure}[h]
\centering
\includegraphics[scale=.9]{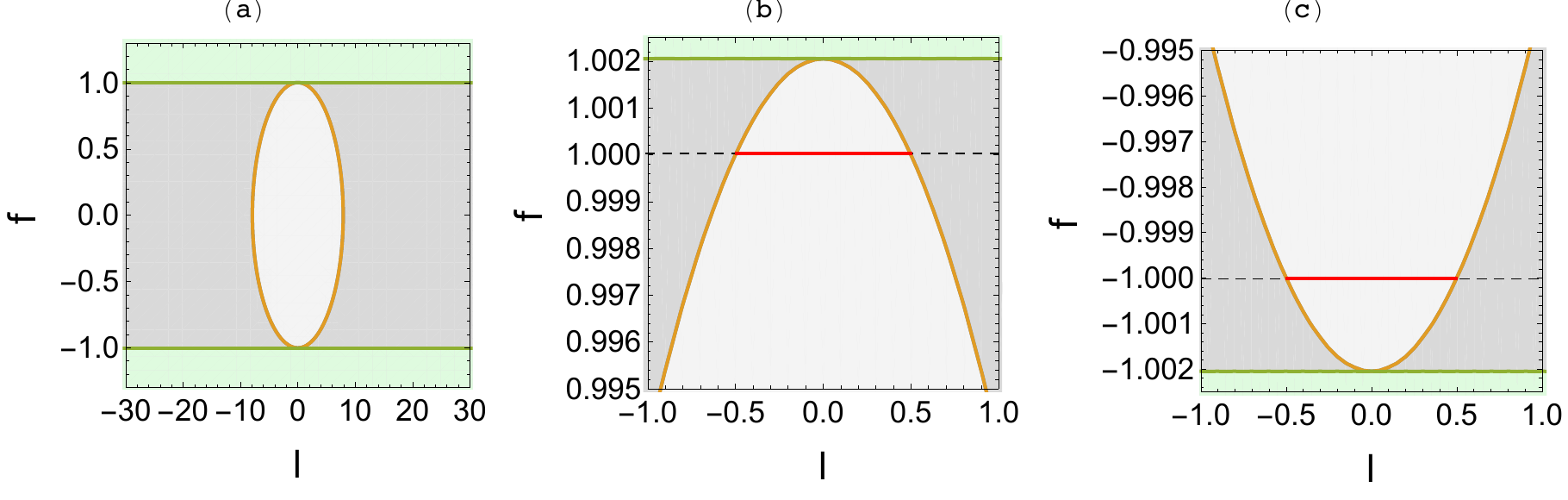}
\caption{(Color online) (a) Stability diagram in the $l\times f$ plane for $U_\mathrm{p}=0.05$, $U_{AB\mathrm{p}}=0.95U_\mathrm{p}$ and $N=10^5$. The dynamically unstable domains are the semi-planes $|f|>\sqrt{p(0)}$ (green area). In turn, the dynamically stable domain is the stripe $|f|<\sqrt{p(0)}$ which is splitted into two disjoint domains, an energetically stable  (light gray area) and an energetically unstable (dark gray area). The energetic phase boundary is a closed curve (orange curve) tangent to the dynamical phase boundaries. The overlap of the SPS and the stability diagram in all $l\times f$ plane excludes the semi-planes $|f|>1$. The energetic phase boundary in the SPS is the orange curve for $\frac{1}{2}<|l|<l_-$ and the red line for $|l|<\frac{1}{2}$. As $\sqrt{p(0)}>1$ the SPS is immersed in the dynamically stable domain, therefore, there is a region of this domain excluded by the overlap which are the stripes $1<f<\sqrt{p(0)}$ and $-\sqrt{p(0)}<f<-1$, see figures (b) and (c). Notice that in our case $\sqrt{p(0)}$ is nearly equal to one (see vertical scale in (b) and (c)).}
\label{Rotulo}
\end{figure}

Since in our discussion we consider only $l\neq 0$, we study the stability in a region of the SPS defined by $|l|>1$. The property that $f_\mathrm{crit}(l)$ is a decreased function of $l$ and that $f\left(\frac{1}{2}\right)=1$ guarantee that one is an upper bound of $f_\mathrm{crit}(l)$. To display the stability diagram in the SPS we need to specify the values of $u_\mathrm{p}$ and $u_{AB\mathrm{p}}$. The strength $u_{ss^\prime}=NU_{ss^\prime}$ is proportional to $N$ which is the total number of atoms in the mixture and $U_{ss^\prime}$ is the interaction strength between atoms of species $s$ and $s^\prime$, in units of $\frac{\hbar^2}{2MR^2}$, independent of $N$. Thus even when the states of the mixture are unchanged we can change the values of $u_{ss^\prime}$ by varying $N$. In Reference [5] they work with a mixture of two hyperfine states of $^{87}\mathrm{Rb}$, the $F=1$, $m_F=1$ and $F=1$, $m_F=0$ states. Since the purpose of the calculations is to illustrate the properties of the QY states discussed in this section, we only demand that the magnitude of the chosen values of the interaction strengths be roughly of the order of the experimental values. We  choose $U_\mathrm{p}=0.05$ and $U_{AB\mathrm{p}}=0.95U_\mathrm{p}$, in units of $\frac{\hbar^2}{2MR^2}$. In Fig. (3) we present the stability diagram in the SPS for different values of $N$. An inspection of these graphs shows that, for fixed values of $N$, $f_\mathrm{crit}(l)$ is a decreased function of $l$ where the size of the window of stability , $0\leq f< f_\mathrm{crit}(l)$, is largest for $l=1$ diminishing when $l$ increases and vanishing at $l=l_\mathrm{crit}$. These graphs also show that $f_\mathrm{crit}(1)$ is weakly dependent on $N$ and nearly equal to one. Concerning $l_\mathrm{crit}$, an exam of these graphs reveals that $l_\mathrm{crit}$ is a increased function of $N$ which has a moderate value equal to $\mathrm{Floor}[7.9]=7$ at $N=10^5$, panel (c), \cite{Floor}. If $N$ decreases, $l_\mathrm{crit}$ also decreases and at $N=10^4$, which is one order of magnitude smaller than the $N=10^5$ mixture, $l_\mathrm{crit}$ is equal to $\mathrm{Floor}[2.5]=2$. This means that at this value of $N$, only QY states with $l=1$ can be energetically stable. Besides, these small values of $l_\mathrm{crit}$ help a possible experimental observation of these effects. If we further lower the value of $N$, we can reach a point where all the energetically stable QY states disappear. This happens when $l_\mathrm{crit}<1$ which requires that $N\leq 1.2\times 10^3$, two orders of magnitude smaller than the $N=10^5$ mixture.

\begin{figure}[h]
\centering
\includegraphics[scale=1]{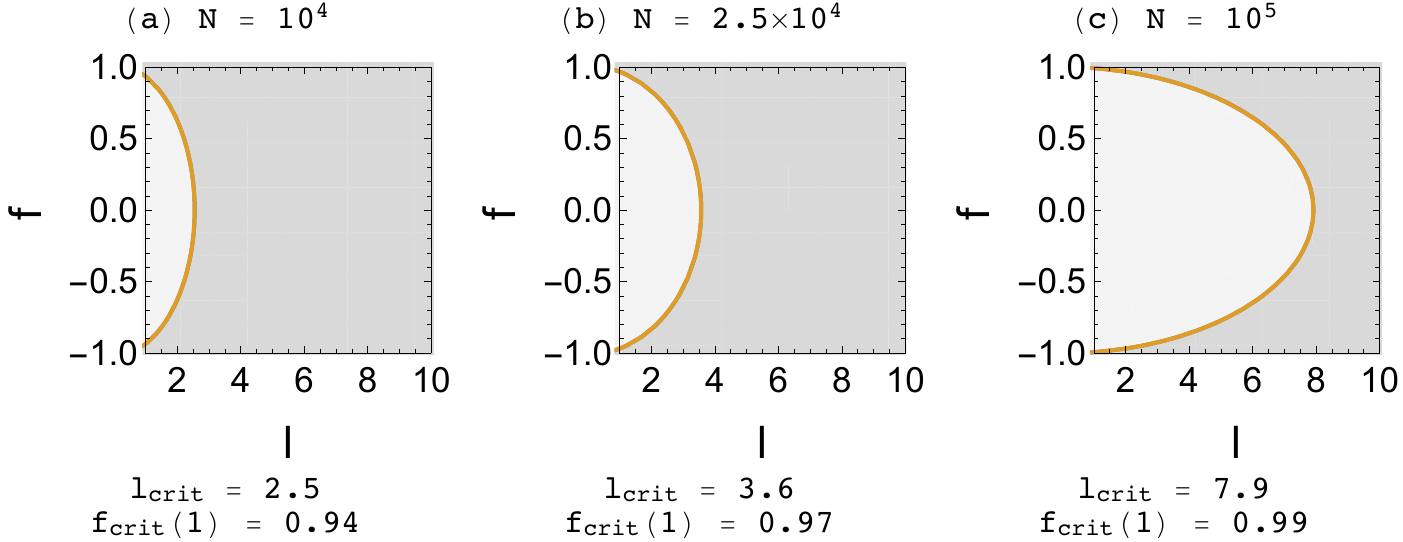}
\caption{(Color online) Stability diagrams in the SPS for $U_\mathrm{p}=0.05$, $U_{AB\mathrm{p}}=0.95U_\mathrm{p}$ and for different values of $N$. The energetically stable QY states are those whose corresponding points in the SPS are localized in the energetically stable domain (light gray area). An inspection of the graphs for fixed values of $N$ shows that $f_\mathrm{crit}(l)$ is a decreased function of $l$ whose largest value occur at $l=1$ vanishing at $l=l_\mathrm{crit}$. In the graphs we see that $f_\mathrm{crit}(1)$ is weakly dependent of $N$ and nearly equal to one. Concerning $l_\mathrm{crit}$, the graphs show that $l_\mathrm{crit}$ is a decreased function of $N$ reaching small values at $N=10^4$ which helps the experimental observation of these effects. If $N$ decreases there is a point where $l_\mathrm{crit}< 1$ and, in this case, all energetically stable QY states disappear.}
\label{Rotulo}
\end{figure}

\section{Summary and conclusions}
In this article we investigate the stability of quantized yrast states in a mixture  of two distinguishable equal mass bosonic atoms confined in a ring. We focus our investigation in the study of the energetic stability of the QY states since the Bloch analysis and the Bogoliubov theory establishes that only energetically stable QY states are capable of sustain a persistent current. 

Based on physical considerations the stability is studied in two different two dimensional planes. One is when we are studying the stability of a single QY state. In this case the study is realized in the $U_{AB}\times U$ plane spanned by the system parameters $U_{AB}$ and $U$ with fixed values of $l$ and $f$ equal to the labels of the QY state. We found that the energetic phase boundary is the positive branch of a hyperbola and the energetically stable domain is the internal region of this positive branch. The stability as function of the interaction strengths follows from the stability diagram in the $U_{AB}\times U$ plane. The measurements to reproduce experimentally the stability diagram in this plane is difficulty by the need to vary the interaction strengths $U_{AB}$ and $U$. However we can think of using the mechanism of Feshbach resonance to overcome this difficult. 

An investigation of the stability of the mixture in the rarefied limit ($f\rightarrow 1$) suggests that we have a mixture of two non mutually interacting gases: the majority and the minority rarefied components. The stability of the mixture is dominated by the stability of the rarefied minority component.

The other is when we are studying the stability at a fixed dynamics. In this case the stability is studied in the $l\times f$ plane spanned by the system parameters $l$ and $f$ with fixed values of the interaction strengths. In this case we extended the investigation to include the study of dynamical stability. In our procedure we begin considering $l$ and $f$ as continuous and unbounded quantities defined in all $l\times f$ plane. We present the stability diagram in this plane which shows that the energetic phase boundary is a closed curve and the energetically stable domain is the internal region of this curve. Until now we consider $l$ and $f$ as continuous and unbounded quantities. However, from the viewpoint of physics, $f$ is a bounded quantity and the correspondence between points in the $l\times f$ plane and QY states  is now restricted to the sector of the $l\times f$ plane of physical significance (SPS) defined by inequalities $-\infty< l<\infty$ and $-1\leq f\leq 1$. What is physically significant is the stability diagram in the SPS determined by the overlap of the stability diagram in all $l\times f$ plane and the SPS. Loosely speaking the stability diagram in the SPS is distilled from the stability diagram in $l\times f$ plane. To reproduce experimentally the stability diagram in the SPS it is necessary to determine the stability of all QY states, which can be done.

At this point we would like to detach interesting properties of the QY states unveiled in our detailed discussion of the properties of the stability diagram in the SPS. They are: (a) All QY states are dynamically stable; (b) Exist a $l_\mathrm{crit}$ in the sense that there is none energetically stable QY state with $l>l_\mathrm{crit}$. In other words, $l_\mathrm{crit}$ is an upper bound of the possible values of $l$ carried by an energetically stable QY state; (c) $l_\mathrm{crit}$ is an increased function of $u_\mathrm{p}-|u_{AB\mathrm{p}}|$. Therefore it decreases when $u_\mathrm{p}-|u_{AB\mathrm{p}}|$ decreases, eventually reaching values of $u_\mathrm{p}-|u_{AB\mathrm{p}}|$ where $l_\mathrm{crit}<1$. This happens when $u_\mathrm{p}-|u_{AB\mathrm{p}}|\leq 3$, in which case all the energetically stable QY states disappear; (d) There is a $f_\mathrm{crit}(l)$ in the sense that, for a given $l$, only QY states with $f$ in the interval $0\leq f< f_\mathrm{crit}(l)$ are energetically stable. As $f_\mathrm{crit}(l)$ is a decreased function of $l$, its largest value occur at $l=1$, diminishing when $l$ increases and vanishing when $l=l_\mathrm{crit}$.

Follows from item (d) that the half-half $f=0$ mixture is the most energetically stable whereas mixtures where one of the components is a rarefied gas ($f\rightarrow 1$) are the most energetically unstable. In other words, small $f$ are fundamentally energetically stable and large $f$ fundamentally energetically unstable.

In our model we are probing the stability of small oscillations of the mixture in the neighborhood of an equilibrium state. In other words, we are probing the stability of its normal modes.  In this case the unstable normal modes are responsible for the emergency of the instability. In the experimental determination of the stability we should follow the time evolution of a state constructed by the action of a weak perturbation of the equilibrium state. Reference [5] is the first experimental work on persistent currents in a mixture. An analysis of the experimental data by our model is beyond its scope since the mechanism which generate the instability in the experiment of Reference [5] is entirely different from the mechanism in our model. 

A final remark. The qualitative aspects of our discussion are generic, that is, independent of the interaction strengths. Of course, this not hold for the quantitative aspects.

\begin{center}
{\bf Acknowledgments}
\end{center}
This work was partially supported by FAPESP under contract 2011/18/998-2.
\appendix


\begin{thebibliography}{10}
\bibitem{Floor} Floor$[x]$ is a function of a continuous variable $x$ defined by Floor$[x]=n$ if $n\leq x< n+1$ with $n$ being the largest integer less or equal to $x$. In our case, this is a prescription to pull out integer values from continuous values of $l$.
\end{thebibliography}
\end{document}